\documentclass[11pt,preprintnumbers,aps,amssymb,nofootinbib,amsmath,superscriptaddress,notitlepage,prd]
{revtex4-1}
\usepackage{epsfig,epsf}
\usepackage{bm} 
\usepackage{color} 
\usepackage{slashed}
\usepackage{relsize}	
\usepackage{soul} 
\usepackage{hyperref}
\usepackage{tensor} 
\newcommand{\beq}{\begin{equation}}
\newcommand{\beql}[1]{\begin{equation}\label{#1}}
\newcommand{\eeq}{\end{equation}}
\def\bal#1\gal{\begin{align}#1\end{align}}
\newcommand{\ball}[1]{\bal\label{#1}}
\newcommand{\eq}[1]{(\ref{#1})}
\newcommand{\fig}[1]{Fig.~\ref{#1}}
\renewcommand{\sec}[1]{Sec.~\ref{#1}}
\renewcommand{\b}[1]{{\bm #1}} 
\newcommand{\unit}[1]{\hat {{\bm #1}}} 

\newcommand{\im}{\,\mathrm{Im}\,}
\newcommand{\real}{\,\mathrm{Re}\,}
\newcommand{\e}{\varepsilon}

\begin{document}

\title{Collisional energy loss and the Chiral Magnetic Effect}

\author{Jeremy Hansen}

\author{Kirill Tuchin}

\affiliation{
Department of Physics and Astronomy, Iowa State University, Ames, Iowa, 50011, USA}

\date{\today}


\begin{abstract}
 
Collisional energy loss of a fast particle in a medium is mostly due to the medium polarization by the electromagnetic fields of the particle. A small fraction of energy is carried away by the Cherenkov radiation. In chiral medium there is an additional contribution to the energy loss due to induction of the anomalous current proportional to the magnetic field. It causes the particle to lose energy mostly in the form of the \emph{chiral}  Cherenkov radiation.  We employ classical electrodynamics, adequate in a wide range of particle energies, to compute the collisional energy loss by a fast particle in a homogenous chiral plasma and apply the results to Quark-Gluon Plasma and a Weyl semimetal. In the later case photon spectrum is strongly enhanced in the ultraviolet and X-ray regions which makes it amenable to experimental investigation.
 Our main observation is that while the collisional energy loss in a non-chiral medium is a slow, at most logarithmic, function of energy $\e$, the chiral Cherenkov radiation is proportional  to $\e^2$ when the recoil is neglected and to $\e$ when it is taken into account.

\end{abstract}

\maketitle

\section{Introduction}\label{sec:i}

A fast charged particle moving through a medium experiences energy loss due to its interaction with medium particles. The collisional part of the energy loss equals the work done by the induced electromagnetic field on displacing the medium particles. If the particle energy is much larger than the typical ionization energy, then the collisional part of its energy loss depends on the macroscopic medium response to the electromagnetic field. The unique feature of such response in materials containing chiral fermions is the Chiral Magnetic Effect \cite{Kharzeev:2004ey,Kharzeev:2007tn,Fukushima:2008xe,Kharzeev:2009fn,Kharzeev:2007jp} which is induction--- by the way of the chiral anomaly \cite{Adler:1969gk,Bell:1969ts}---of anomalous electric current flowing in the magnetic field direction.  The work spent by the particle on inducing the anomalous current contributes to its energy loss. The goal of this paper is to calculate this anomalous part of  the collisional energy loss. 

 The classical calculation  of the collisional energy loss in the non-chiral medium was first preformed by Fermi \cite{Fermi:1940zz}.  We generalize his calculation to the chiral medium with anomalous response to the magnetic field, utilizing the recently derived expressions for the electromagnetic field \cite{Tuchin:2020pbg}. Our main conclusion is that the energy loss in the chiral medium is proportional to the particle energy itself and comes mostly in the form of the chiral Cherenkov radiation---an analogue of the Cherenkov radiation that exist only in the chiral medium. This is in striking contrast with the collisional energy loss in non-chiral medium  which is independent of the particle's energy in the ultra-relativistic limit. We also argue that in a wide range of particle energies, quantum corrections due to the recoil effects are small.  

 The collisional energy loss spectrum is given by Eqs.~\eq{b2}--\eq{b14}. It contains the anomalous contribution, mostly due to the chiral Cherenkov radiation, which is clearly seen in \fig{fig:qgp} for Quark-Gluon Plasma and in \fig{fig:semimetal} for a Weyl semimetal. In the later case the photon spectrum is strongly enhanced in the ultraviolet and X-ray regions which makes it amenable to experimental investigation.


\section{Electromagnetic fields in chiral medium}\label{sec:a}

Electrodynamics of isotropic chiral medium is characterized by the emergence of the anomalous current proportional to the magnetic field viz.\ $\b j_A = \sigma_\chi \b B$, where $\sigma_\chi$ is the chiral conductivity \cite{Kharzeev:2009pj,Fukushima:2008xe}. As a result, the field equations for a point charge $q$ moving in the positive $z$ direction with constant velocity $v$ read:
\begin{subequations}\label{A1}
\bal
&\b \nabla \times \b B = \partial_t \b D +\sigma_\chi \b B+ qv\unit z \delta(z-vt)\delta (\b b) \,,\label{a1}\\
&\b \nabla\cdot \b D= q \delta(z-vt)\delta (\b b)\,,\label{a2}\\
&\b \nabla \times \b E =-\partial_t \b B\,,\label{a3}\\
&\b \nabla\cdot \b B=0\,,\label{a4}
\gal 
\end{subequations}
where $\b b$ denotes the transverse components of the position vector $\b r$. 
The solution to \eq{A1} with $\b D_\omega =\epsilon (\omega)\b E_\omega$, where $E_z= \frac{1}{2\pi}\int_{-\infty}^{\infty} E_{z\omega}e^{-i\omega t}d\omega$ etc., was derived in \cite{Tuchin:2020pbg} as a superposition of the helicity states $\lambda=\pm 1$, which are the eigenstates of the curl operator in the Cartesian coordinates:
\begin{subequations}\label{p4}
\bal
  \b B(\b r,t)&= \int \frac{d^2k_\bot d\omega}{(2\pi)^3}e^{i\b k\cdot \b r-i\omega t}\sum_\lambda\b\epsilon_{\lambda \b k} \frac{q \unit z \cdot\b \epsilon_{\lambda\b k}^*\lambda k}{k_\bot^2+\omega^2(1/v^2-\epsilon)-\lambda \sigma_\chi k }\,,\label{p6}
 \\
  \b E(\b r,t)&= \int \frac{d^2k_\bot d\omega}{(2\pi)^3}e^{i\b k\cdot \b r-i\omega t}\left(  \sum_\lambda\b\epsilon_{\lambda \b k}\frac{iq\omega \unit z \cdot\b \epsilon_{\lambda\b k}^* }{k_\bot^2+\omega^2(1/v^2-\epsilon)-\lambda \sigma_\chi k }+ \unit k \frac{q}{ivk\e}\right)\,, \label{p7}
 \gal
 \end{subequations}
 where $\b k = \b k_\bot + (\omega/v)\unit z$ is the wave vector,   $k= \sqrt{k_\bot^2+\omega^2/v^2}$ its length and  $\b\epsilon_{\lambda \b k}$ are the circular polarization vectors satisfying the conditions  $\b\epsilon_{\lambda \b k}\cdot\b\epsilon_{\mu \b k}^*=\delta_{\lambda\mu}$, $\b\epsilon_{\lambda \b k}\cdot \b k=0$ and the identity
\ball{p8}
 i\unit k \times\b \epsilon_{\lambda \b k }= \lambda\b \epsilon_{\lambda \b k }\,.
 \gal
Summations over $\lambda$ are performed using the polarization sums given in Appendix of \cite{Tuchin:2020pbg}. The space-time dependence of the electromagnetic field given by the  integrals \eq{p6},\eq{p7} was approximately evaluated in \cite{Tuchin:2014iua,Li:2016tel,Tuchin:2020pbg} assuming the low frequency limit of a conductor $\epsilon = 1+i\sigma/\omega$ and used to compute the effect of the chiral anomaly on the magnetic field produced in relativistic heavy-ion collisions. 
 
To compute the energy loss we need only the frequency components of the fields, see \eq{b2}. These can be computed exactly. For illustration, consider 
\ball{p10} 
 \b B_\omega(\b r)=& \int \frac{d^2k_\bot}{(2\pi)^2}\frac{q\, k \, e^{i\omega z/v+i\b k_\bot \cdot \b b}}{[k_\bot^2+\omega^2(1/v^2-\epsilon)]^2-(\sigma_\chi k)^2}\nonumber\\
& \times 
 \left\{ [k_\bot^2+\omega^2(1/v^2-\epsilon)] \sum_\lambda\lambda \b\epsilon_{\lambda \b k}(\unit z \cdot\b \epsilon_{\lambda\b k}^* )+\sigma_\chi k \sum_\lambda\b\epsilon_{\lambda \b k}(\unit z \cdot\b \epsilon_{\lambda\b k}^* )\right\}
 \gal
Its azimuthal component is 
\ball{p12}
B_{\phi\omega}(\b r)= & \int \frac{d^2k_\bot}{(2\pi)^2}\frac{q\, k \, e^{i\omega z/v+i\b k_\bot \cdot \b b}}{[k_\bot^2+\omega^2(1/v^2-\epsilon)]^2-(\sigma_\chi k)^2}\nonumber\\
& \times 
 \left\{ [k_\bot^2+\omega^2(1/v^2-\epsilon)] \frac{-ik_\bot}{k}\cos\theta+\sigma_\chi k \frac{-k_zk_\bot}{k^2}\sin\theta\right\}\,,
 \gal
where $\b k_\bot \cdot \b b= k_\bot b \cos\theta$. Integration over $\theta$ and then over $k_\bot$ yields \eq{a11} below. Other field components an be obtained in a similar way with the following result:
\begin{subequations}\label{A10}
\bal
B_{\phi\omega}(\b r)=& \frac{q}{2\pi}\frac{e^{i\omega z/v}}{k_1^2-k_2^2}\sum_{\nu = 1}^2(-1)^{\nu+1} k_\nu(k_\nu ^2-s^2)K_1(bk_\nu)\,,\label{a11}\\
B_{b\omega}(\b r)=& \sigma_\chi\frac{q}{2\pi}\frac{i\omega }{v}\frac{e^{i\omega z/v}}{k_1^2-k_2^2}\sum_{\nu = 1}^2(-1)^\nu k_\nu K_1(bk_\nu)\,,\label{a12}\\
B_{z\omega}(\b r)=& \sigma_\chi\frac{q}{2\pi}\frac{e^{i\omega z/v}}{k_1^2-k_2^2}\sum_{\nu = 1}^2 (-1)^{\nu+1} k^2_\nu K_0(bk_\nu )\,,\label{a13}\\
E_{z\omega}(\b r)=& \frac{q}{2\pi}\frac{i\omega }{v^2\epsilon}\frac{e^{i\omega z/v}}{k_1^2-k_2^2}\sum_{\nu = 1}^2(-1)^{\nu+1}  \left[(v^2\epsilon-1)(k_\nu^2-s^2)-\sigma_\chi^2\right]K_0(bk_\nu)\,,\label{a14}\\
E_{b\omega}(\b r)=& \frac{q}{2\pi}\frac{1}{v\epsilon}\frac{e^{i\omega z/v}}{k_1^2-k_2^2}\sum_{\nu =1}^2(-1)^{\nu+1}  k_\nu\left(k_\nu^2-s^2-\sigma_\chi^2\right)K_1(bk_\nu)\,,\label{a15}\\
E_{\phi\omega}(\b r)=&vB_{b\omega}(\b r)\,,\label{a16}
\gal
\end{subequations}
 where 
 \ball{a20}
 k_\nu^2= s^2-\frac{\sigma_\chi^2}{2}+(-1)^\nu \sigma_\chi \sqrt{\omega^2\epsilon+\frac{\sigma^2_\chi}{4}}
 \gal
with $\nu=1,2$ and
\ball{a21}
s^2=\omega^2\left(\frac{1}{v^2}-\epsilon(\omega)\right)\,.
\gal
Without loss of generality we assume that $\sigma_\chi>0$ which implies that  $k_2^2>k_1^2$. 
The plasma permittivity is well described by
\ball{a23}
\epsilon =1-\frac{\omega_p^2}{\omega^2+i\omega\Gamma}\,,
\gal
 where $\omega_p$ is the plasma frequency and the damping constant $\Gamma$ is related to the electrical conductivity.

\section{Collisional energy loss}\label{sec:b}

The energy loss rate can be computed as the flux of the Poynting vector out of a cylinder of radius $a$ coaxial with the particle path. For a particle moving with velocity $v$ along the $z$-axis the total loss per unit length reads
\ball{b2} 
-\frac{d\e}{dz}= 2\pi a\int_{-\infty}^{\infty}(E_\phi B_z- E_zB_\phi)dt= 2a\real\int_{0}^{\infty}(E_{\phi\omega} B^*_{z\omega}- E_{z\omega}B^*_{\phi\omega})d\omega\,.
\gal
To calculate the integral over $\omega$  we first isolate the contribution of the pole in $1/\epsilon$ at $\omega=\omega_p$ using the rule 
\ball{b4}
\frac{1}{\epsilon} = \frac{\omega^2}{\omega^2-\omega_p^2+i0}= -i\pi\omega_p^2 \delta(\omega^2-\omega_p^2) + \mathcal{P}\frac{\omega^2}{\omega^2-\omega_p^2}\,,
\gal
 where it is assumed that $\Gamma\ll \omega_p$.
Substituting the field components from \eq{A10} into \eq{b2} and replacing $1/\epsilon$ by its imaginary part one derives 
\ball{b6}
-\frac{d\e^\text{pole}}{dz}= \frac{q^2\omega_p^2}{4\pi v^2}K_0\left(a\omega_p/v\right)\real\left\{
a\sqrt{\omega_p^2/v^2-\sigma_\chi^2}K_1\left(a\sqrt{\omega_p^2/v^2-\sigma_\chi^2}\right)\right\}\,.
\gal

Away from the pole, the permittivity is real. In this case the contribution to the integral over $\omega$ comes from those domains of $\omega$ where at least one of $k_\nu$'s is imaginary. There are two such domains (A) and (B).
Domain (A) $k_1^2<k_2^2<0$. Inspection of \eq{a20} reveals that $k_2^2<0$ if either $\omega^2>\omega^2_+$ or $\omega^2<\omega^2_-$ where 
 \ball{b8}
 \omega_\pm^2=\frac{-2(1/v^2-1)\omega_p^2+\sigma_\chi^2/v^2\pm 
 \sqrt{[2(1/v^2-1)\omega_p^2-\sigma_\chi^2/v^2]^2-4(1/v^2-1)^2\omega_p^4}}
 {2(1/v^2-1)^2}\,.
 \gal
 Additionally, if $\omega_p<\sigma_\chi/\sqrt{2}$ the inequality 
 \ball{b9}
 0<\omega<\sqrt{\frac{\sigma_\chi^2/2-\omega_p^2}{1/v^2-1}}\,
 \gal
must be satisfied.  Domain (B) $k_1^2<0<k_2^2$ corresponds to in the interval $\omega_-^2<\omega^2<\omega_+^2$. We note that in vacuum\footnote{The term `vacuum' refers to Eqs.~\eq{A1} with $\epsilon=1$ and finite $\sigma_\chi$, which is a version of axion electrodynamics \cite{Carroll:1989vb}.}, i.e. when $\omega_p=0$ domain (A) is empty because \eq{b8} and \eq{b9} are never satisfied at any $\sigma_\chi$. In contrast, when $\sigma_\chi=0$, domain (B) is empty because then $k_1=k_2$. 
 
It can be readily verified that $\im k_1<0$  at all $\omega$'s, while  $\im k_2<0$ when $\epsilon>0$ and positive otherwise. It is not difficult to compute the real part of the Poynting vector in both cases. However, 
later we will be primarily concerned with the ultrarelativistic limit $v\to 1$ in which the energy loss is dominated by  high frequencies $\omega\gg \omega_p$. In this case $\epsilon>0$ and the imaginary parts of $k_{1,2}$ are negative.  It is convenient to denote $k_+^2= -k_1^2$ and $k_-^2= -k_2^2$, so that 
 in  domain (A) $\im k_{1,2}= k_\pm <0$ while in domain (B) $\im k_1= k_+<0$. We derive
\bal
\real(E_{\phi\omega}B_{z\omega}^*)\Big|_{(A)}= &\frac{q^2}{(2\pi)^2}\frac{\sigma_\chi^2\omega}{ (k_+^2-k_-^2)^2}
\Big\{\frac{\pi}{2a}(k_+^2+k_-^2)   \nonumber \\
& -\frac{\pi^2}{4}k_+k_-^2\left[ J_1(ak_+)Y_0(ak_-)-J_0(ak_-)Y_1(ak_+)\right]
\nonumber \\
&
-\frac{\pi^2}{4}k_+^2k_-\left[ J_1(ak_-)Y_0(ak_+)-J_0(ak_+)Y_1(ak_-)\right]
\Big\} \,,\label{b10}\\
\real(E_{\phi\omega}B_{z\omega}^*)\Big|_{(B)}= &\frac{q^2}{(2\pi)^2}\frac{\sigma_\chi^2\omega}{ (k_+^2-k_-^2)^2}
\Big\{\frac{\pi}{2a}k_+^2 \nonumber\\
&-\frac{\pi}{2}k_2^2k_+J_1(ak_+)K_0(ak_2)+\frac{\pi}{2}k_+^2k_2K_1(ak_2)J_0(ak_+)\Big\}\,, \label{b11}
\gal
\bal
&\real(E_{z\omega}B_{\phi\omega}^*)\Big|_{(A)}\nonumber\\
&= \frac{q^2}{(2\pi)^2}\frac{\omega}{v^2\epsilon (k_+^2-k_-^2)^2}
\Big\{-\frac{\pi}{2a}\left[(v^2\epsilon-1)(k_+^2+s^2)+\sigma_\chi^2\right](k_+^2+s^2) \nonumber\\
&-\frac{\pi}{2a}\left[(v^2\epsilon-1)(k_-^2+s^2)+\sigma_\chi^2\right](k_-^2+s^2) \nonumber\\
&-\frac{\pi^2}{4}\left[(v^2\epsilon-1)(k_+^2+s^2)+\sigma_\chi^2\right](k_-^2+s^2)k_-\left[ J_0(ak_+)Y_1(ak_-)-J_1(ak_-)Y_0(ak_+)\right] \nonumber\\
&-\frac{\pi^2}{4}\left[(v^2\epsilon-1)(k_-^2+s^2)+\sigma_\chi^2\right](k_+^2+s^2)k_+\left[ J_0(ak_-)Y_1(ak_+)-J_1(ak_+)Y_0(ak_-)\right]\Big\} \,,\label{b13}\\
&\real(E_{z\omega}B_{\phi\omega}^*)\Big|_{(B)}\nonumber\\
&= \frac{q^2}{(2\pi)^2}\frac{\omega}{v^2\epsilon (k_+^2-k_-^2)^2}
\Big\{-\frac{\pi}{2a}\left[(v^2\epsilon-1)(k_+^2+s^2)+\sigma_\chi^2\right](k_+^2+s^2) \nonumber\\
&-\frac{\pi}{2}\left[(v^2\epsilon-1)(k_+^2+s^2)+\sigma_\chi^2\right](s^2-k_2^2)k_2J_0(ak_+)K_1(ak_2)\nonumber\\
&-\frac{\pi}{2}\left[(v^2\epsilon-1)(s^2-k_2^2)+\sigma_\chi^2\right](k_+^2+s^2)k_+J_1(ak_+)K_0(ak_2)\Big\}\,.
\label{b14}
\gal
Substituting these equations into \eq{b2} we obtain frequency spectrum of the energy loss  $-d\e/dzd\omega$ which is plotted in \fig{fig:qgp} and \fig{fig:semimetal}. The $\omega$-integral can be computed exactly in several important approximations that we now consider.

\subsection{Ultrarelativistic limit}\label{sec:b2}

In the limit $ak_\nu\ll 1$, the contribution of the pole \eq{b6} is proportional to large logarithm $\ln a$, the term \eq{b14} is independent of $a$,  whereas the remaining terms are suppressed by the positive powers of $ak_\nu$. The corresponding energy loss reads
\ball{b20}
-\frac{d\e}{dz}= \frac{q^2}{4\pi}\frac{\omega_p^2}{v^2}\ln \frac{1.12 v}{a\omega_p}- \frac{q^2}{4\pi}\int_{k_1^2<0<k_2^2}\frac{\omega}{v^2\epsilon}\frac{(s^2-k_1^2)(v^2\epsilon-1)+\sigma_\chi^2}{k_1^2-k_2^2}d\omega\,.
\gal
The integration domain (B) simplifies in the ultrarelativistic limit $v\to 1$: $\omega_p^2/\sigma_\chi<\omega< \gamma^2\sigma_\chi$, where $\gamma=(1-v^2)^{-1/2}$. Expanding the integrand at large frequencies, assuming $\omega\gg \sigma_\chi$, yields
\ball{b22}
- \frac{q^2}{4\pi}\frac{1}{2\sigma_\chi}\int^{\gamma^2\sigma_\chi}
\left(- \frac{\sigma_\chi \omega}{\gamma^2}+\sigma_\chi^2\right) d\omega\,,
\gal
where the precise value of the lower limit is irrelevant as long as $\gamma\gg 1$. Integrating one obtains
\ball{b25}
-\frac{d\e}{dz}= \frac{q^2}{4\pi v^2}\left( \omega_p^2\ln \frac{v}{a\omega_p}+ \frac{1}{4}\gamma^2\sigma_\chi^2\right)\,.
\gal
We observe that the energy loss due to the anomaly, represented by the second term in \eq{b25}, dominates at high energies. Inclusion of quantum effects produces the logarithmic dependence of the first term on $\gamma$  but this does not change our conclusion.

\subsection{Non-chiral medium}\label{sec:b3}

In the limit $\sigma_\chi\to 0$ the contributions of \eq{b10},\eq{b11} and \eq{b14} vanish. The finite limit  emerges from \eq{b13} which along with \eq{b6} yields
\ball{b30}
-\frac{d\e}{dz}= \frac{q^2}{4\pi}\frac{\omega_p^2}{v^2}K_0(a\omega_p/v)(a\omega_p/v) K_1(a\omega_p/v)
+\frac{q^2}{4\pi v^2}\int_{s^2<0} \omega \left( v^2-\frac{1}{\epsilon}\right)d\omega \,.
\gal
The second term vanishes in plasma since $\epsilon<1$ implies that $s^2$ is always positive, see  \eq{a21} and \eq{a23}. However, if  medium contains bound states, then the second term contributes when the velocity of the particle is larger than the phase velocity of light in the medium. A single bound state of frequency $\omega_0$ contributes to the permittivity as 
\ball{b31}
\epsilon(\omega) = 1-\frac{\omega_p^2}{\omega^2-\omega_0^2+i\omega \Gamma}
\gal
In this case \eq{b30} is generalized as
\ball{b33}
-\frac{d\e}{dz}=& \frac{q^2}{4\pi}\frac{\omega_p^2}{v^2}K_0\left(a\sqrt{\omega_p^2+\omega_0^2}/v\right)\left(a\sqrt{\omega_p^2+\omega_0^2}/v\right) K_1\left(a\sqrt{\omega_p^2+\omega_0^2}/v\right)\nonumber\\
&+\frac{q^2}{4\pi v^2}\int_{s^2<0} \omega \left( v^2-\frac{1}{\epsilon}\right)d\omega 
\gal
Neglecting $\Gamma$, the integration region $s^2<0$ is equivalent to  $(1-\epsilon(0)v^2)/(1-v^2)<\omega^2/\omega_0^2<1$ if $v<1/\sqrt{\epsilon(0)}$ and to $\omega<\omega_0$ if $v>1/\sqrt{\epsilon(0)}$. Integration over $\omega$ in the second term yields the well-known Fermi's result \cite{Fermi:1940zz}.

%
\subsection{Cherenkov radiation}\label{sec:k}

Some of the collisional energy loss emerges in the form of the Cherenkov radiation. In the non-chiral medium it is included in the second term in \eq{b30} (provided that $\epsilon(0)$ is finite, as explained in the previous subsection) and is small compared to the large first term that describes excitation of the longitudinal oscillations in the medium (medium polarization). 

In the chiral medium the Cherenkov radiation emerges even when $\epsilon=1$, which is known as the chiral (or, in a different context, vacuum) Cherenkov radiation \cite{Tuchin:2018sqe,Huang:2018hgk,Lehnert:2004hq,Lehnert:2004be}.\footnote{It was proposed to be a test of the Lorentz symmetry violation in \cite{Carroll:1989vb,Lehnert:2004hq,Lehnert:2004be,Klinkhamer:2004hg,Mattingly:2005re,Kostelecky:2002ue,Jacobson:2005bg,Altschul:2006zz,Altschul:2007kr,Nascimento:2007rb}.} It is generated by the anomalous electromagnetic current in the presence of the moving charged particle. The Cherenkov radiation is that part of the total energy flux moving radially away from the particle which is finite at $a\to \infty$. It can be computed by replacing the Bessel functions appearing in \eq{b10}--\eq{b14} with their asymptotic expressions. In particular, the rate of the chiral Cherenkov radiation emitted in a unit interval of frequencies by an ultrarelativistic particle in empty space ($\epsilon=1$) is given by 
\ball{k2}
\frac{dW}{d\omega} =-\frac{d\e}{dz\omega d\omega}\Big|_{a\to \infty}=
\frac{q^2}{4\pi} \left\{ \frac{1}{2}\left( 1-\frac{1}{v^2}\right)+\frac{\sigma_\chi}{2\omega}+\frac{(1+v^2)\sigma_\chi^2}{8v^2\omega^2}+\ldots\right\}\,, \quad \omega< \sigma_\chi\gamma^2\,.
\gal 
which comes about from \eq{b14}. Expansion in powers of $\sigma_\chi/\omega$ is justified  for the ultrarelativistic particle. Eq.~\eq{k2} is derived neglecting the fermion recoil which proportional to $\hbar \omega$. It is a good approximation as long as $\omega_+= \sigma_\chi\gamma^2\ll \e$, in other words when $\gamma\ll m/\sigma_\chi$,  where $m$ is the particle mass. 
The total radiated power $P$ is obtained by integrating \eq{k2} over $\omega d\omega$. It is dominated by the upper limit so that only the first two terms contribute at large $\gamma$ with the result:
\ball{k4}
P= \frac{q^2}{4\pi}\frac{\sigma_\chi^2\gamma^2}{4}\,.
\gal
 We observe that the spectrum \eq{k2} is exactly the same as \eq{b22} which indicates that all energy lost by the ultrarelativistic particle due to the anomalous current is radiated as the chiral Cherenkov radiation.

The spectrum of the chiral Cherenkov radiation was previously computed by one of the authors in the leading order of QED with the result \cite{Tuchin:2018sqe}
\ball{d55}
\frac{dW^\text{quant}}{d\omega}&=  \frac{q^2}{(4\pi) 2\omega }\left\{ \sigma_\chi \left(\frac{x^2}{2}-x+1\right)-\frac{m^2}{\e}x\right\}\,,\quad \omega< \omega_M\,,
\gal
where $x= \omega/\e$ is the fraction of the fermion energy  carried away by the radiated photon and 
\ball{d56}
\omega_M=\frac{\e}{1+m^2/(\sigma_\chi\e)}\,.
\gal
Photon spectrum always extends all the way to $\omega_M$ since $\omega_M <\sigma_\chi\gamma^2$. Moreover, since $\omega_M< \e$ and hence $x<1$, \eq{d55} is valid even at $\gamma\gg m/\sigma_\chi$, in contrast to the classical formula \eq{k2}. The classical limit of \eq{d55}  is recovered in the limit $x\ll 1$: the term in \eq{d55} proportional to $\sigma_\chi$ reduces  to the second term in \eq{k2}, while the second term in \eq{d55} reduces to the first term in \eq{k2}. 
The total radiation power is
\ball{d60}
P^\text{quant}= \frac{q^2}{4\pi}\frac{\sigma_\chi \e}{3}\,,
\gal
where the terms of order $m/\e = 1/\gamma$ were neglected.  Evidently, the effect of the recoil on the energy loss is to reduce the energy dependence from $\e^2$ to $\e$. 

\section{Discussion}\label{sec:s}


The classical calculation performed in this paper captures the main feature of the energy loss in chiral medium, namely, its  much faster increase with the particle energy that in a non-chiral medium. Taking the recoil effects into account, the energy loss is proportional to energy $\e$. In contrast, energy dependence of the collisional energy loss in non-chiral medium is at most logarithmic. The conventional radiative energy loss is likewise proportional to energy in the non-coherent Bethe-Heitler  (BH) regime. The ratio of the energy loss due to the chiral Cherenkov ($\chi C$) effect to the conventional radiative loss is 
\ball{s1}
\frac{\Delta \e^{\chi C}}{\Delta \e^{BH}}\sim \frac{\sigma_\chi}{e^2 T}\sim \frac{\mu_5}{T}\,,
\gal
 where $T$ is the plasma temperature and $\mu_5$ is the axial chemical potential. The coherence effects reduce the energy dependence of the radiative energy loss  to $\sqrt{\e}$ (see review \cite{Peigne:2008wu}). This significantly increases the ratio \eq{s1}. In this paper we assumed that distribution of the topological charge density is homogenous and therefore there are no coherence effects on the chiral Cherenkov radiation. This is a good approximation as long as the coherence length associated with photon radiation is smaller than the distance over which the topological charge density significantly varies.  This maybe the case in the nuclear matter where there is evidence---supported by the theoretical arguments \cite{Zhitnitsky:2012ej,Zhitnitsky:2013hs}---of the topological domains of nearly constant density in wide range of temperatures \cite{Horvath:2003yj,Horvath:2005rv,Horvath:2005cv,Alexandru:2005bn,Ilgenfritz:2007xu,Ilgenfritz:2008ia,Kovalenko:2004xm,Bruckmann:2011ve}. 

\begin{figure}[t]
      \includegraphics[height=6cm]{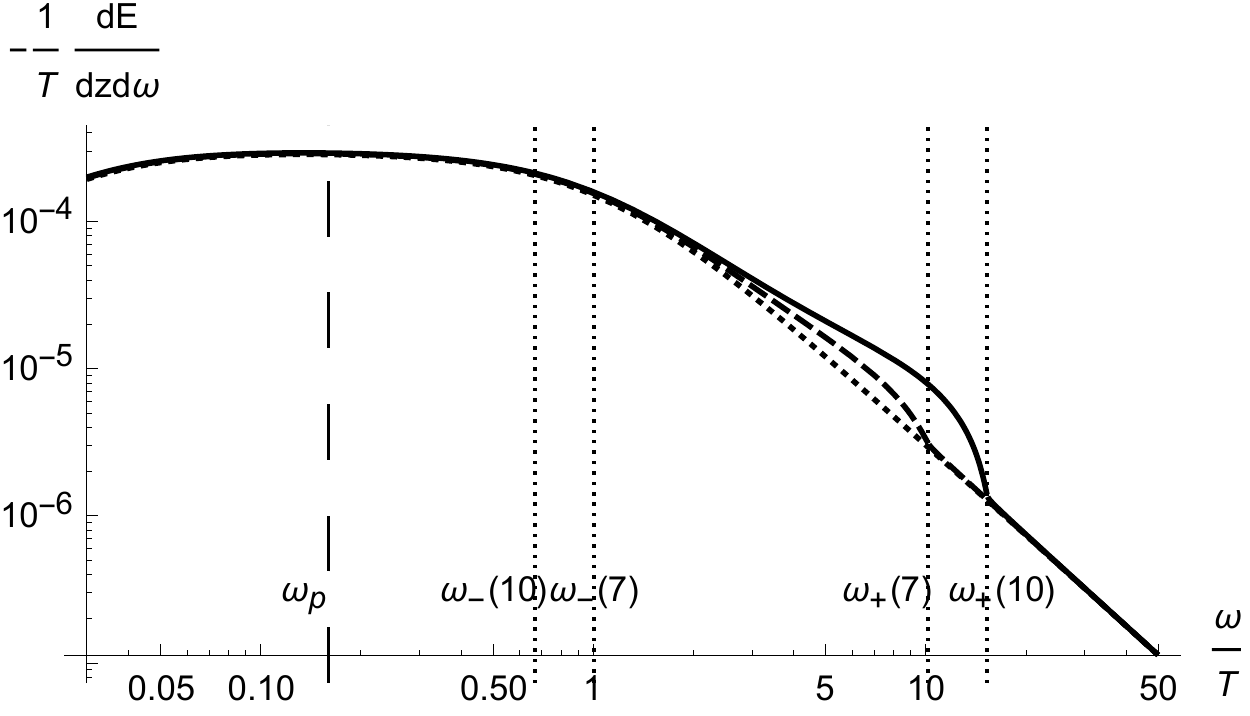} 
  \caption{Electromagnetic part of the collisional energy loss spectrum of  a $d$-quark with $\gamma=20$ in Quark-Gluon Plasma. Plasma parameters: $\omega_p= 0.16T$, $\Gamma= 1.11T$  \cite{Ding:2010ga}, $m=T=250$~MeV. Solid line: $\sigma_\chi=10$~MeV, dashed line: $\sigma_\chi=7$~MeV, dotted line: $\sigma_\chi=0$. $\omega_\pm$ are defined in \eq{b8}. }
\label{fig:qgp}
\end{figure}

\fig{fig:qgp} displays the spectrum of the collisional energy loss by a fast particle in Quark-Gluon Plasma computed using the results of \sec{sec:b}. We emphasize that this is a purely electromagnetic effect. The chiral Cherenkov radiation emerges as a bump between $\omega_-$ and $\omega_+$. The relevant parameters are inferred from the lattice calculations \cite{Ding:2010ga} or by the way of educated guess in the case of the chiral conductivity. The quantum corrections due to the recoil would shift the UV endpoint of the anomalous contribution to the left since $\omega_M< \omega_+=\sigma_\chi\gamma^2$. However, the overall effect of the recoil  is not that significant since $\omega_M=0.44 \e$ is not too close to unity. The rate of energy loss due to the anomaly is about $10^{-4}$ of particle energy per unit fm regardless of particle energy as indicated by \eq{k4}.  This is of course much smaller than the QCD energy loss mechanisms \cite{dEnterria:2009xfs} at not too high energies; however the QCD fraction decreases as $1/\sqrt{\e}$.

\begin{figure}[t]
      \includegraphics[height=6cm]{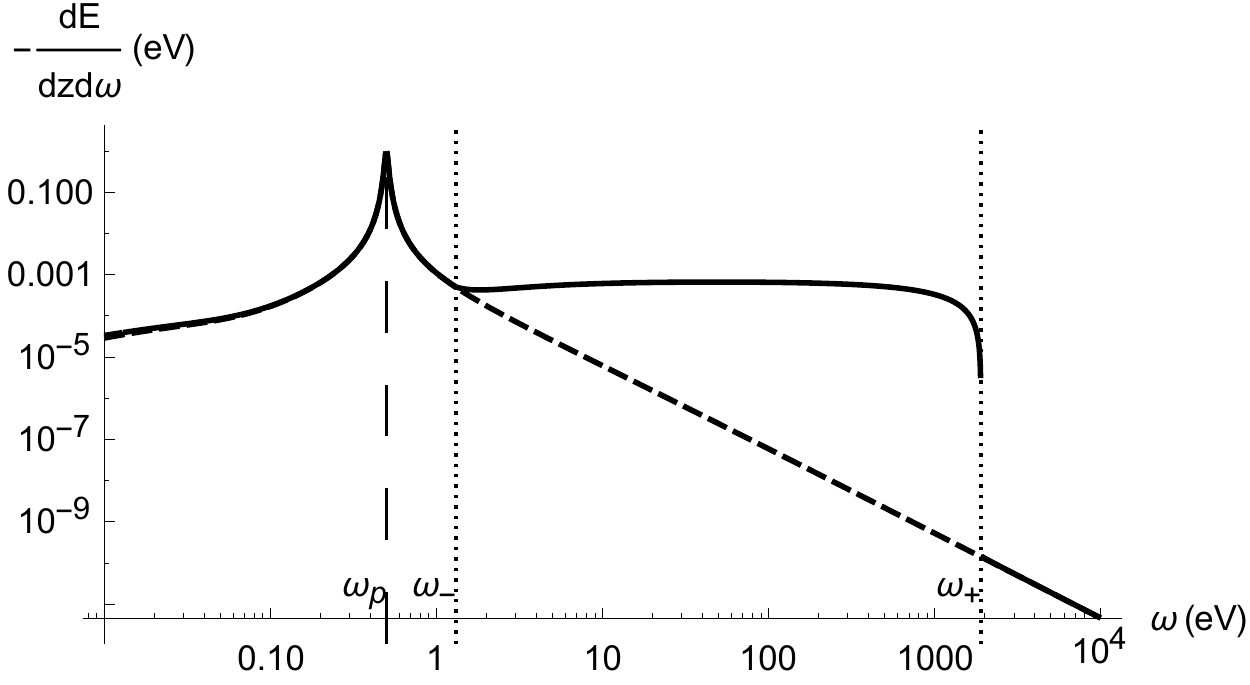} 
  \caption{Collisional energy loss spectrum of electron  with $\gamma=100$ in a semimetal with parameters $\omega_p= 0.5$~eV, $\Gamma=0.025$~eV (so that its conductivity is 10 eV at room tempearture)  \cite{Huang:2015eia} and $m=0.5$~MeV. Solid line: $\sigma_\chi=0.19$~eV \cite{Xu:2015cga,Lv:2015pya}, dashed line:  $\sigma_\chi=0$. $\omega_\pm$ are defined in \eq{b8}. The seeming discontinuity at $\omega=\omega_+$ is a visual artifact.  }
\label{fig:semimetal}
\end{figure}

In anisotropic chiral medium such as Weyl semimetals there is an additional anomalous current $\b j_{AH}=\b b\times \b E$ that generates the anomalous Hall effect  \cite{Zyuzin:2012tv,Grushin:2012mt,Klinkhamer:2004hg}. Parameter $\b b$ is the distance between the Weyl nodes in the momentum space (not to be confused with the impact parameter used in \sec{sec:a}). The spectrum of the corresponding chiral Cherenkov radiation was computed in \cite{Tuchin:2018mte}. In the ultra-relativistic limit, the energy loss equals the total radiated power and is given by \eq{d60} with $\sigma_\chi$ replaced by $b$ (assuming that electron's velocity is parallel to $\b b$). To estimate the energy loss  in a semimetal reported in \cite{Xu:2015cga,Lv:2015pya} we use $b=(\alpha/\pi)80$~eV. According to \eq{s1} at room temperatures most of energy is lost due to chiral Cherenkov radiation. The energy loss spectrum for a typical semimetal computed using the results of \sec{sec:b} is displayed in \fig{fig:semimetal}. The recoil effect is negligible since $\omega_M\lesssim \sigma_\chi\gamma^2\ll \e$. One observes significant enhancement of the ultraviolet and X-ray regions of the photon spectrum which presents an exciting opportunity for experimental study of the chiral anomaly effects. 

\acknowledgments
This work  was supported in part by the U.S. Department of Energy under Grant No.\ DE-FG02-87ER40371.



\end{document}